# Features of Excess Conductivity Behavior in a Magnetic Superconductor $Dy_{0.6}Y_{0.4}Rh_{3.85}Ru_{0.15}B_4$
Low Temperature Physics

A. L. Solovjov, A. V. Terekhov, E. V. Petrenko, L. V. Omelchenko, and Zhang Cuiping

View Online 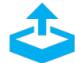 Export Citation 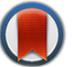 CrossMark

**ARTICLES YOU MAY BE INTERESTED IN**

Sub-kelvin Andreev reflection spectroscopy of superconducting gaps in FeSe
Low Temperature Physics **45**, 1222 (2019); https://doi.org/10.1063/10.0000133

Atomic structures and nanoscale electronic states on the surface of $MgB_2$ superconductor observed by scanning tunneling microscopy and spectroscopy
Low Temperature Physics **45**, 1209 (2019); https://doi.org/10.1063/10.0000132

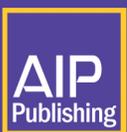

# Features of Excess Conductivity Behavior in a Magnetic Superconductor $Dy_{0.6}Y_{0.4}Rh_{3.85}Ru_{0.15}B_4$

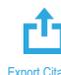

View Online    Export Citation    CrossMark

A. L. Solovjov,[1,a)] A. V. Terekhov,[1] E. V. Petrenko,[1] L. V. Omelchenko,[1] and Zhang Cuiping[2]

**AFFILIATIONS**

[1]B. Verkin Institute for Low Temperature Physics and Engineering, National Academy of Sciences of Ukraine, 47 Nauki Ave., Kharkov 61103, Ukraine
[2]Superconducting Material Research Center (SMRC), Northwest Institute for Non-Ferrous Metal Research (NIN), Xi'an, China

[a)]Email: solovjov@ilt.kharkov.ua

**ABSTRACT**

The temperature dependencies of the excess conductivity $\sigma'(T)$ and possible pseudogap (PG), $\Delta^*(T)$, in a $Dy_{0.6}Y_{0.4}Rh_{3.85}Ru_{0.15}B_4$ polycrystal were studied for the first time. It was shown that $\sigma'(T)$ near $T_c$ is well described by the Aslamazov–Larkin (AL) fluctuation theory, demonstrating a 3D–2D crossover with increasing temperature. Using the crossover temperature $T_0$, the coherence length along the c axis, $\xi_c(0)$, was determined. Above the level of $T_{2D} > T_0$, an unusual dependence $\sigma'(T)$ was found, which is not described by the fluctuation theories in the range from $T_0$ to $T_{FM}$, at which a ferromagnetic transition occurs. The range in which superconducting fluctuations exist is apparently quite narrow and amounts to $\Delta T_{fl} \approx 2.8$ K. The resulting temperature dependence of the PG parameter $\Delta^*(T)$ has the form typical of magnetic superconductors with features at $T_{max} \approx 154$ K and the temperature of a possible structural transition at $T_s \sim 95$ K. Below $T_s$, dependence $\Delta^*(T)$ has a shape typical for PG in cuprates, which suggests that the PG state can be realized in $Dy_{0.6}Y_{0.4}Rh_{3.85}Ru_{0.15}B_4$ in this temperature range. Comparison of $\Delta^*(T)$ with the Peters–Bauer theory made it possible to determine the density of local pairs near $T_c$, $\langle n_\uparrow n_\downarrow \rangle(TG) \approx 0.35$, which is 1.17 times greater than in optimally doped $YBa_2Cu_3O_{7-\delta}$ single crystals.

Published under license by AIP Publishing. https://doi.org/10.1063/10.0000125

## 1. INTRODUCTION

Recent studies of physical properties of new materials have shown that researchers increasingly encounter so-called unconventional superconductivity.[1,2] For example, excitonic or magnonic mechanisms of superconducting (SC) pairing of charge carriers in such materials can be different from those of phonons.[3] In addition, the pairing symmetry in unconventional superconductors may differ from that described by the Bardeen–Cooper–Schrieffer (BCS) theory, and the superconducting order parameter often vanishes at some points of the momentum space (e.g. in the case of p- or d-wave symmetry).[3] In the BCS theory, the total spin of an electron pair is equal to zero ($S = 0$), whereas, for example, in triplet superconductors, $S = 1$, which also falls beyond the scope of this theory. Nonconventional superconductors additionally include those in which magnetism coexists with superconductivity (magnetic superconductors), which also contradicts the BCS theory.[3–6]

One of the most prominent representatives of magnetic superconductors are the triple rare-earth rhodium borides $RERh_4B_4$ (RE is a rare-earth element).[6] In these materials, various types of magnetic ordering can be observed depending on the type of rare earth (such as ferromagnetic (FM), antiferromagnetic (AFM), as well as spiral spatially modulated magnetic structures). In the case of FM superconductors (e.g. $ErRh_4B_4$), transition to the superconducting state occurs at higher temperatures, followed by FM ordering at lower temperatures, which suppresses superconductivity.

This is observed in the study of certain bulk properties (magnetization, electrical resistance, etc.), e.g. in the form of recursive superconductivity (i.e. the transition of a material from a superconducting to a normal state at low temperatures under the action of internal magnetism.[4,6] In the case of AFM materials, such as $NdRh_4B_4$, $SmRh_4B_4$, $TmRh_4B_4$, the AFM transition was also observed below the temperature of the SC transition, but, in contrast to FM compounds, superconductivity was suppressed only partially, thus providing the coexistence of these two types of ordering down to the lowest temperatures.[4,6]

The most interesting case of superconductivity-magnetism coexistence was observed in systems where magnetic rare earth was partially replaced by a non-magnetic element.[7] Such compounds include, among others, rare-earth rhodium borides $Dy_{1-x}Y_xRh_4B_4$ ($x = 0, 0.2, 0.4$) with a tetragonal body-centered crystal structure of the LuRu$_4$B$_4$ type.[6] Magnetic ordering in these materials appears above the SC transition temperature and coexists with superconductivity to the lowest temperatures.[8,9]

In Ref. 8, it was shown that the magnetic transition in Dy$_{1-x}$Y$_x$Rh$_4$B$_4$ with $x = 0$, 0.2, 0.4 is ferrimagnetic, and the magnetic transition temperature $T_C$ strongly depends on the concentration of nonmagnetic Y and decreases with increasing concentration from 37 K in DyRh DyRh$_4$B$_4$ to 7 K in Dy$_{0.2}$Y$_{0.8}$Rh$_4$B$_4$. Consequently, when concentration of Y increases from 4.7 K for DyRh$_4$B$_4$ to 10.5 K in YRh$_4$B$_4$, the superconducting transition temperature $T_c$ also increases.[8] Measurements of heat capacity of Dy$_{0.8}$Y$_{0.2}$Rh$_4$B$_4$, Dy$_{0.6}$Y$_{0.4}$Rh$_4$B$_4$ and Dy$_{0.6}$Y$_{0.4}$Rh$_{3.85}$Ru$_{0.15}$B$_4$ showed that a further magnetic transformation can occur below the superconducting transition temperature.[10]

It is not improbable that low-temperature magnetic transitions are possible at other concentrations of Y. In particular, it was found recently that the behavior of certain physical quantities in magnetic superconductors Dy$_{1-x}$Y$_x$Rh$_4$B$_4$ ($x = 0$, 0.2, 0.4) is uncharacteristic of systems with conventional superconductivity. These specifics include the paramagnetic Meissner effect[11,12] and the nonmonotonic behavior of dependencies $H_{c2}(T)$ and $\Delta(T)$.[9,13–15]

Studies of solid solutions of Dy(Rh$_{1-x}$Ru$_x$)$_4$B$_4$ showed that the replacement of rhodium by ruthenium may change the type of magnetic interactions: AFM ordering for $x < 0.5$ and ferromagnetic ordering for $x > 0.5$. This may be due to the change in the Ruderman–Kittel–Kasuya–Yosida (RKKY)-exchange interaction that occurs between Dy atoms through conduction electrons of Rh or Ru atoms.[16] We have recently investigated the magnetic properties of Dy$_{0.6}$Y$_{0.4}$Rh$_{3.85}$Ru$_{0.15}$B$_4$ (to be published soon) and showed that below 19 K, there is a transition to the FM state ($\mu_{sat} \approx 6.2\mu_B$ per Dy$^{3+}$ ion at 2 K), and below 6.7 K, superconductivity takes place and both of these states coexist.

Thus, the study of physical properties of the Dy$_{1-y}$Y$_y$ (Rh, Ru)$_4$B$_4$ boride family with different contents of dysprosium (responsible for magnetic interactions) and rhodium-ruthenium compound (responsible for both magnetic interactions and superconductivity) is of considerable interest in terms of studying various aspects of superconductivity-magnetism coexistence, as well as for revealing signs of non-conventional superconductivity. In this work, for the first time, we thoroughly investigated the behavior of excess conductivity of Dy$_{0.6}$Y$_{0.4}$Rh$_{3.85}$Ru$_{0.15}$B$_4$ near $T_c$ within the framework of existing fluctuation theories, and addressed the issue of the possible existence of a pseudogap state, its nature, and susceptibility to magnetic ordering.

### 1.1. Samples and Experimental Methods

The samples of Dy$_{0.6}$Y$_{0.4}$Rh$_{3.85}$Ru$_{0.15}$B$_4$ were prepared by argon-arc melting of the initial components, followed by annealing for several days, as described in Ref 14. As a result, we obtained single-phase textured polycrystalline samples with a LuRu$_4$B$_4$ structure (space group $I4/mmm$) (Fig. 1), as evidenced by the results of X-ray phase-sensitive and X-ray diffraction analyses.[8,9] The critical temperature of the SC transition is $T_c$ ($R = 0$) $\sim$6.4 K (Fig. 2). Based on literature sources, we believe that the geometric parameters of the crystal lattice in our case are: $a = b \approx 7.45$ Å, $c \approx 15$ Å.[9] Partial replacement of Rh with Ru made it possible to synthesize samples at normal pressure, which would be impossible without such replacement.[6] It is known that the tetrahedra of Rh$_4$B$_4$/RU$_4$B$_4$ have different orientations and are shown enlarged in Fig. 1. In the LURU$_4$B$_4$ structure, the Dy and Y atoms occupy the Lu positions. It can be seen that the Dy atoms in the planes are surrounded by nonequivalent tetrahedra of Rh$_4$B$_4$/Ru$_4$B$_4$, because the distances between Rh or Ru atoms in diversely oriented tetrahedra are noticeably different: 2.98 and 3.10 Å, respectively.

Electrical resistance measurements were performed using a standard four-probe circuit on a Quantum Design PPMS-9 automated system at alternating current $I = 8$ mA $f = 97$ Hz). Figure 2 shows the temperature dependence of the resistivity $\rho(T)$ of the test sample. In the temperature range between $T^* = (167 \pm 0.5)$ K and ~280 K, the dependence $\rho(T)$ is linear with the slope $a = d\rho/dT = 0.14$. The slope was determined by linear curve fitting and confirms the excellent linearity of $\rho(T)$ with a standard error of $0.0012 \pm 0.0002$ over the entire temperature range noted. As usual, $T^* \gg T_c$ was defined as the temperature at which the resistive curve deviates from linearity to lower values[5,17] (Fig. 2). It can be seen that below $T^*$ $\rho(T)$ takes the form characteristic of magnetic superconductors with positive curvature.[5,18]

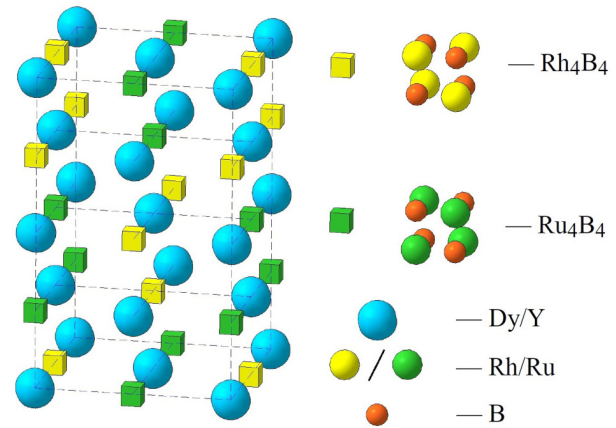

**FIG. 1.** Idealized tetragonal body-centered crystalline structure of Dy$_{0.6}$Y$_{0.4}$Rh$_{3.85}$Ru$_{0.15}$B$_4$.

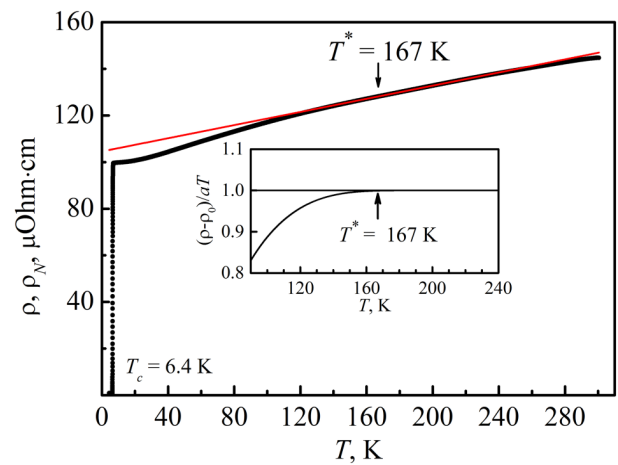

**FIG. 2.** Temperature dependence $\rho(T)$ of a Dy$_{0.6}$Y$_{0.4}$Rh$_{3.85}$Ru$_{0.15}$B$_4$ polycrystal. The straight red line defines $\rho_N(T)$ which is extrapolated to the lower temperatures. The insert shows a more accurate method for determining $T^* = 167$ K using the criterion $(\rho - \rho_0)/aT = 1$.[19]

For a more accurate determination of $T^*$ (with an accuracy of $\pm 0.5$ K), we used a modified straight line equation $[\rho(T) - \rho_0]/aT = 1$,[19] as shown in the insert in Fig. 2. Here, the same as above, $a = d\rho/dT$ denotes the slope of the temperature dependence of resistivity in the normal state, $\rho_N(T)$, which is extrapolated to the low-temperature region, and $\rho_0$ is the residual resistance determined by the intersection of $\rho_N$ with the Y axis.

Both methods give the same values of $T^*$.

From resistive measurements, the fluctuation contributions to the excess conductivity $\sigma'(T)$ were determined and the temperature dependence of the pseudogap parameter $\Delta^*(T)$ was calculated and analyzed. The results obtained show that in the region of SC fluctuations near $T_c$ $\sigma'(T)$ is well approximated by the Aslamazov–Larkin (AL) fluctuation theory for three-dimensional systems.[20] However, the SC fluctuation region is very small and unexpectedly expands with increasing temperature $\sigma'(T)$, showing its maximum near the FM transition temperature $T_{FM} \sim 19$ K. The corresponding dependence $\Delta^*(T)$ has a form similar to that found in FeSe$_{0.94}$ polycrystals.[21] Despite this, the density of nearby local pairs $\langle n_\uparrow n_\downarrow \rangle$, derived from the comparison of $\Delta^*(T)$ with the Peters–Bauer theory[22] turned out to be 1.17 times greater. A detailed analysis of these results is provided below.

## 2. RESULTS
### 2.1. Fluctuation conductivity

The temperature dependence of excess conductivity was determined conventionally using the equation in Refs. 17 and 23

$$\sigma'(T) = \sigma(T) - \sigma_N(T) = \frac{1}{\rho(T)} - \frac{1}{\rho_N(T)}. \quad (1)$$

An important parameter for further analysis is the reduced temperature

$$\varepsilon = \frac{T - T_c^{mf}}{T_c^{mf}}, \quad (2)$$

which is included in all equations in this article. Here, $T_c^{mf} > T_c$ is the critical temperature in the mean-field approximation, which separates the region of fluctuation conductivity (FLC) from the region of critical fluctuations or fluctuations of the SC parameter of the Δ order immediately near $T_c$, unaccounted for in the Ginzburg–Landau theory.[24,25] This indicates that the correct determination of $T_c^{mf}$ plays a decisive role in the FLC and PG calculations. For this purpose, we use the fact[5,17] that near $T_c$ in all HTSCs, $\sigma'(T)$ is always described by the standard equation of the Aslamazov–Larkin theory[20] with a critical exponent $\lambda = -1/2$, which determines FLC in any three-dimensional (3D) system:

$$\sigma'_{AL3D} = C_{3D} \frac{e^2}{32\hbar\xi_c(0)} \varepsilon^{-1/2}, \quad (3)$$

where $\xi_c(0)$ is the coherence length along the $c$ axis and $C_{3D}$ is the coefficient (C-factor) by which the theory data defined by Eq. (3) should be multiplied to align them with experimental results. As is known,[26,32] the closer the C-factor to 1, the better the structure of the sample, and vice versa. Trimerization of a HTSC near $T_c$ is most likely caused by Gaussian fluctuations in 2D metals that include HTSC compounds exhibiting a pronounced quasi-two-dimensional anisotropy of the conductive properties [Ref. 17 and references therein].

Taking account of Gaussian fluctuations brings the critical temperature of an ideal 2D metal to zero (Mermin–Wagner–Hohenberg theorem), and its final value is only obtained when trimerization factors are included.[27–29] Thus, 3D FLC is always realized in HTSCs when $T$ approaches $T_c$.[30,31] As a result, it is determined by extrapolating the linear dependence $\sigma'^{-2}$ in the 3D-fluctuation region from $T$ to its intersection with the temperature axis, because, when $T \to T_c^{mf}$, $\sigma'$ should diverge as $(T - T_c^{mf})$ [see Eq. (3)].[32] Note that, in each case, $T_c^{mf} > T_c$. In cuprates, this shift is ~1–2 K, which, to a first approximation, gives the value of critical fluctuations above $T_c$.

It should be emphasized that, when $T_c^{mf}$ is determined correctly, $\sigma'(T)$ in the 3D fluctuation region near $T_c$ is always described by Eq. (3). Another characteristic temperature is the Ginzburg temperature $T_G > T_c^{mf}$, marked as $\ln \varepsilon_G = -5.0$ in Fig. 3, up to which fluctuation theories are applicable. This temperature is usually determined by the Ginzburg criterion, which applies when the GL mean-field theory becomes inapplicable when describing the SC transition.[33,34] It can be seen (Fig. 3) that below $T_G$, the data deviate downward from the AL line, indicating a transition to critical fluctuations near $T_c$.[32,35]

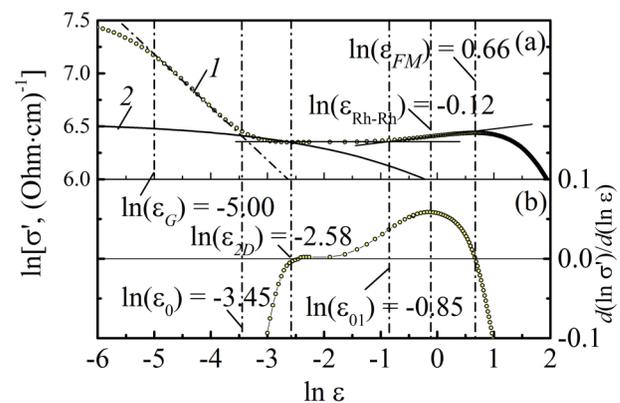

FIG. 3. (a) Dependence of $\ln \sigma'$ on $\ln \varepsilon$ of a Dy$_{0.6}$Y$_{0.4}$Rh$_{3.85}$Ru$_{0.15}$B$_4$ polycrystal in comparison with fluctuation theories near $T_c$: 3D AL (1, dotted line), 2D MT (2, solid curve). (b) Derivative $d(\ln \sigma')/d(\ln \varepsilon)$ of $\ln \varepsilon$. All characteristic temperatures are indicated by vertical dashed lines.

By determining $T_c^{mf} = 6.62$ K, Eq. (2) can be used to find $\varepsilon(T)$ and construct the dependence $\sigma'(\varepsilon)$ in the double logarithmic coordinates accepted in the literature (Fig. 3). Within the model of local pairs (LP),[36–38] it was shown that the FCL measured for all HTSCs without exception always shows a crossover from a 3D state ($\xi_c(T) > d$) near $T_c$ to 2D ($\xi_c(T) < d$) as $T$ increases [Refs. 5, 17, 31, 32, and references therein], where $d = c \approx 15$ Å is the cell dimension along the $c$ axis.[30] At the crossover temperature $T_0$ $\xi_c(T_0) = \xi_c(0) \varepsilon_0^{-1/2} = d$.[30] Hence,

$$\xi_c(0) = d\sqrt{\varepsilon_0}, \quad (4)$$

which makes it possible to determine $\xi_c(0)$. On the upper panel of Fig. 3, the dependence of $\ln \sigma'$ on $\ln \varepsilon$ is constructed in comparison with fluctuation theories. As expected, above $T_G \approx 6.67$ K ($\ln \varepsilon_G = -5.0$) and up to $T_0 = 6.8$ K ($\ln \varepsilon_0 = -3.45$), $\sigma'(T)$ is well described by Eq. (3) (dashed line $1$) with $\xi_c(0) = (2.67 \pm 0.02)$ Å determined according to Eq. (4), and $C_{3D} = 0.38$. Above $T_0$, the data deviate upward from the linear dependence, indicating a transition to the region of 2D fluctuations. Clearly, between $T_0$ and $T_{2D} = 7.1$ K ($\ln \varepsilon_{2D} = -2.58$), $\ln \sigma'(\ln \varepsilon)$ is determined by the Maki–Thompson fluctuation contribution (MT2D)[39,40] [Eq. (5)] (solid curve $2$) of the Hikami–Larkin (HL) theory for HTSCs:

$$\sigma'_{MT2D} = \frac{e^2}{8d\hbar} \frac{1}{1 - \alpha/\delta} \ln\left(\frac{\delta}{\alpha} \frac{1 + \alpha + \sqrt{1 + 2\alpha}}{1 + \delta + \sqrt{1 + 2\delta}}\right) \varepsilon^{-1}, \quad (5)$$

applied in the field of 2D fluctuations.[26,30] Here, the communication parameter

$$\alpha = 2\left(\frac{\xi_c(0)}{d}\right)^2 \varepsilon^{-1}, \quad (6)$$

steaming parameter

$$\delta = \beta \frac{16}{\pi\hbar} \left(\frac{\xi_c(0)}{d}\right)^2 k_B T \tau_\varphi, \quad (7)$$

and the phase relaxation time, $\tau_\varphi$ is given by equation

$$\tau_\varphi \beta T = \pi\hbar/(8k_B\varepsilon) = A/\varepsilon, \quad (8)$$

where $A = 2.998 \cdot 10^{-12}$ s·K. Factor $\beta = 1.203$ $(l/\xi_{ab})$, where $l$ is the mean free path and $\xi_{ab}$ is the coherence length in the $ab$ plane, takes into account the approximation of the pure limit ($l > \xi$).[17,26] In this case, however, the region of MT fluctuations is very small, $\Delta T_{fl} = T_{2D} - T_G \approx 0.4$ K [Fig. 3(a)]. At $\ln \varepsilon_{2D} = -2.58$, which we designated as $T_{2D}$, the experimental points deviate upward from the MT curve, and the first derivative of the experimental curve becomes zero [Fig. 3(b)]. Above $T_{2D}$, FLC no longer conforms to the classical fluctuation theories.

In HTSCs, with an increase in temperature above the region of 2D fluctuations, the experimental data normally deviate downward from the MT curve.[17,26] The unusual behavior of FLC discovered in this case is most likely due to the presence, as noted above, of a large magnetic moment in dysprosium as part of our compound ($\sim 6.2\mu_B$). As a result, the $\ln \sigma'$–$\ln \varepsilon$ relationship in the indicated temperature range has several singular points. It can be seen [Fig. 3(a)] that above $T_{2D}$ the experimental data can be approximated by two lines that intersect at $\ln \varepsilon_{01} \approx -0.85$, showing a dramatic change in the dependence slope at this temperature. At this point, the first derivative has an inflection point [Fig. 3(b)], which is confirmed by the second derivative (not shown) demonstrating a maximum at this point.

It must be emphasized that at this temperature, temperature dependence $\Delta^*(T)$ displays a small but sharp minimum (Fig. 5) designated as $T_{01}$. This minimum at $\Delta^*(T)$ is observed in all the studied HTSCs: cuprates,[5,17,41] pnictides[18] and chalcogenides of FeSe.[21] It corresponds to the temperature $T_{01}$ that places an upper limit on the region of SC fluctuations near $T_c$, where fluctuation Cooper pairs (FCPs) behave almost like classical Cooper pairs, but without long-range ordering – the so-called 'short-range phase correlations'.[22,42–45] Moreover, in this temperature range, the $\ln \sigma'$–$\ln \varepsilon$ dependence always conforms to the classical fluctuation theories of AL[20] and MT.[30]

Based on these considerations, we believe that in the case of $Dy_{0.6}Y_{0.4}Rh_{3.85}Ru_{0.15}B_4$, this minimum also corresponds to the temperature $T_{01} \approx 9.4$ K, which is designated in Fig. 3(a) as $\ln \varepsilon_{01}$. Accordingly, in $Dy_{0.6}Y_{0.4}Rh_{3.85}Ru_{0.15}B_4$, the SC fluctuation range is very small: $\Delta T_{fl} = T_{01} - T_G = (9.4-6.67)$ K $\approx 2.8$ K. This is notably smaller than $\Delta T_{fl} = 10.4$ K obtained for the $FeSe_{0.94}$ sample with $T_c = 9$ K and without defects,[21] but, curiously, greater than $\Delta T_{fl} = 1.45$ K, measured for an optimally subsidized (OS) single crystal of $YBa_2Cu_3O_{7-\delta}$ (YBCO) with $T_c \sim 91.1$ K.[46] This result indicates that the sample under study may contain a certain number of defects, presumably in the form of grain boundaries forming a polycrystal.

In the local pairs model, it is assumed that in HTSC, $\xi_c(T) = \xi_c(0) (T/T_c^{mf} - 1)^{-1/2} = \xi_c(0) \varepsilon^{-1/2}$,[47] increasing with decreasing temperature, at $T = T_{01}$ becomes equal to the distance between the conducting layers $d_{01}$ (in YBCO, these are $CuO_2$ planes) and connects them with the Josephson interaction,[31] which explains the appearance of 2D FLC below $T_{01}$.[17,26] Accordingly, $\xi_c(T) = d$ at $T = T_0$, and below $T_0$ in HTSCs, 3D FLC is implemented, as noted above. Since $\xi_c(0) = (2.67 \pm 0.02)$ Å is already defined above according to (4), the simple relation $\xi_c(0) = d\varepsilon_0^{1/2} = d_{01}\varepsilon_{01}^{1/2}$ makes it possible to find $d_{01} = d(\varepsilon_0/\varepsilon_{01})^{1/2} \approx 4.08$ Å, taking into account that in this case $d = 15$ Å. In fact, this is the distance between the Dy/Y atoms and the Rh/Ru/B tetrahedra, and therefore between the corresponding conducting planes in $Dy_{0.6}Y_{0.4}Rh_{3.85}Ru_{0.15}B_4$ along the $c$ axis (Fig. 1). Indeed, $4d_{01} \approx 16.3$ Å is in good agreement with the unit cell size along the $c$ axis.

Above $T_{01}$ ($\ln \varepsilon_{01} \approx -0.85$), FLC increases rapidly, reaching a maximum at the Curie temperature, $T_{FM} \sim 19$ K, obtained from magnetic measurements. Accordingly, at this temperature, the first derivative is equal to zero [Fig. 3(b)]. Between $T_{FM}$ and $T_{01}$, there is another singular point, which is the inflection point on the $\ln \sigma'$–$\ln \varepsilon$ dependence at $T = T_{Rh–Rh}$, which is almost invisible on the scale used, but is observed as a maximum on the first derivative at $\ln \varepsilon_{Rh-Rh} \approx -0.12$ [Fig. 3(b)]. It is of interest to estimate to what characteristic distances this temperature corresponds in the structure of $Dy_{0.6}Y_{0.4}Rh_{3.85}Ru_{0.15}B_4$. At $T_{Rh-Rh}$, we obtain $d_{Rh-Rh} = d(\varepsilon_0/\varepsilon_{Rh-Rh})^{1/2} \approx 2.85$ Å, which, possibly accidentally, is the distance between Rh atoms (or, respectively, Ru atoms) in $Rh/Ru-B_4$ tetrahedra designated, respectively, as the yellow and green cubes in Fig. 1. Parameters of the sample are presented in Table I.

It can be assumed that at $T < T_{FM}$, ordered magnetic moments begin to prevent the formation of FLC more intensively. This process slows down significantly at $T \leq T_{01}$, indicating the increasing role of SC fluctuations in the FLC formation. Curiously, according to our estimates, $d_{01} \approx 4.08$ Å $= d/4$. This result suggests that the forming quasicoherent FCPs restore the effective distance between the conducting layers to its geometric value. Below $T_{2D}$ [ln $\varepsilon_{2D} = -2.6$ in Fig. 3(a)], a rapid increase in FLC begins, which becomes very intense in the region of 3D fluctuations at $T < T_0$ [ln $\varepsilon_0 = -3.45$ in Fig. 3(a)].

In all probability, this is not only due to the rapid increase in the number of FCPs, but also to a sharp increase in the superfluid density $\rho_s$ in the region of 3D fluctuations,[44,48–50] since near $T_c$, FCPs are already covered by the Josephson interaction in the entire bulk of the superconductor.[17,26]

We can thus assume that it is the interplay of magnetism and superconductivity that is responsible for the unusual dependence of ln $\sigma'$ on ln $\varepsilon$ discovered in Dy$_{0.6}$Y$_{0.4}$Rh$_{3.85}$Ru$_{0.15}$B$_4$. It should be expected that the dependence $\Delta^*(T)$, which is analyzed in the next section, should also be unusual.

### 2.2. Analysis of Dependence $\Delta^*(T)$

In resistive measurements of HTSC cuprates, the pseudogap is manifested as the deviation of the longitudinal resistivity $\rho(T)$ at $T < T^*$ from its linear dependence in the normal condition above $T^*$.[23] This gives rise to excess conductivity $\sigma'(T)$ (1). It is assumed that, if there were no processes in the HTSCs causing the PG to open at $T^*$, then $\rho(T)$ would remain linear up to $\sim T_c$. It is thus obvious that the excess conductivity $\sigma'(T)$ appears as a result of the PG opening and, therefore, should contain information about its magnitude and temperature dependence.

We also share the view that the PG in cuprates arises due to the formation of local pairs (LPs) at $T < T^*$.[17,41–44] In this case, the classical fluctuation theories of both AL and MT, which is modified for HTSCs by Hikami and Larkin (HL),[30] well describe the experimental dependence $\sigma'(T)$ in cuprates, but only up to $T_{01}$, i.e. usually in the range of $\sim 15$ K above $T_c$.[5,17] Understandably, to obtain information about $\Delta^*(T)$, we need an equation that would describe the entire experimental curve from $T^*$ to $T_c$ and contain $\Delta^*(T)$ in an explicit form. In the absence of a rigorous theory, this equation was proposed in:[17,41]

$$\sigma'(\varepsilon) = \frac{e^2 A_4 (1 - T/T^*)(\exp(-\Delta^*/T))}{16\hbar\xi_c(0)\sqrt{2\varepsilon_{c0}^*}\mathrm{sh}(2\varepsilon/\varepsilon_{c0}^*)}, \quad (9)$$

where $(1 - T/T^*)$ and $\exp(-\Delta^*/T)$ take into account, respectively, the dynamics of LP formation at $T \leq T^*$ and their destruction near $T_c$; $A_4$ is a numerical coefficient that has the meaning of the C-factor in the theory of FLC.[17,26,32] Parameters $T^*$, $\varepsilon$ and $\xi_c(0)$ are defined from the analysis of resistivity and FLC. It is important that other parameters, such as the theoretical parameter $\varepsilon_{c0}^*$,[51] coefficient $A_4$, and $\Delta^*(T_G)$, can also be defined from experiment within the framework of the LP model.

It must be emphasized that in HTSC cuprates, at $T \leq T^*$, not only do all the parameters of the samples change, but the density of electronic states (DOS) at the Fermi level also begins to decrease,[52,53] which, by definition, is called the pseudogap.[54] As may be assumed, this also involves the rearrangement of the Fermi surface,[23,55] which breaks up into Fermi arcs below $T^*$.[50,53] It is believed that a correct understanding of the PG physics should also answer the question about the mechanism of SC pairing in HTSCs, which remains controversial.[17,22] However, we do not know of any DOS measurements for Dy$_{0.6}$Y$_{0.4}$Rh$_{3.85}$Ru$_{0.15}$B$_4$. Therefore, the question of the PG appearance in this system remains open. Let us analyze $\sigma'(T)$ in Dy$_{0.6}$Y$_{0.4}$Rh$_{3.85}$Ru$_{0.15}$B$_4$ in the framework of our LP model using Eqs. (9) and (10), but without referring to $\Delta^*(T)$ as the pseudogap.

Analysis of the ln $\sigma'$–ln $\varepsilon$ dependence (Fig. 4) shows that in the temperature range of 41 K < $T$ < 71 K, indicated by arrows at ln $\varepsilon_{c01} = 1.64$ and ln $\varepsilon_{c02} = 2.27$, $\sigma'^{-1} \sim \exp \varepsilon$.[51] This peculiarity turns out to be one of the main properties of the majority of HTSCs.[5,17] As a result, in the range of $\varepsilon_{c01} < \varepsilon < \varepsilon_{c02}$ (insert in Fig. 4), ln ($\sigma'^{-1}$) is a linear function of $\varepsilon$ with a slope $\alpha^* = 0.14$, which defines the parameter $\varepsilon_{c0}^* = 1/\alpha^* \approx 7.14$.[51] This makes it possible to obtain reliable values of $\varepsilon_{c0}^*$, which, as established,[5,17,41] significantly affect the form of the theoretical curves shown in Fig. 4 at $T \gg T_{01}$. Accordingly, in order to find the coefficient $A_4$, calculations will be performed using Eq. (9) and combined with the experiment in the 3D AL fluctuation region near Tc, where it is a linear function of the reduced temperature $\varepsilon$, with a slope of $\lambda = -1/2$[17,41] (Fig. 4). As seen in Fig. 4 and Eq. (9) with $A_4 = 11$, $\varepsilon_{c0}^* = 7.14$ and $\Delta^*(T_G) = 3.5k_B T_c$ (red curve in Fig. 4), as expected, well describes the experiment in the temperature range of $T^*$ to $T_G$. An exception is the temperature range from $T_{FM}$ to $T_0$, where, as noted above, a strong influence of magnetism is assumed. Curiously, in this temperature range with $T$ exceeding ln $\varepsilon \approx -1.4$, theoretical curve (9) increases rapidly and, starting from ln $(\varepsilon_{FM}) = 0.66$, describes the experiment perfectly well.

The correct value of $\Delta^*(T_G)$, which is used in Eq. (9), is found by combining the theory with experimental points constructed as ln $\sigma'$ from $1/T$, as e.g. in Refs. 5, 41, and 46. In addition, it is assumed that $\Delta^*(T_G) = \Delta$ (0), where $\Delta$ is the SC gap.[48,56] We emphasize that it is the magnitude of $\Delta^*(T_G)$ that determines the true value of the PG and is used to estimate the BCS ratio

**TABLE I.** Values of the parameters describing the specifics of $\sigma'(T)$ in Dy$_{0.6}$Y$_{0.4}$Rh$_{3.85}$Ru$_{0.15}$B$_4$.

| $\rho(10\ K)$, μOhm·cm | $T_c$, K | $T_c^{mf}$, K | $T_g$, K | $T_0$, K | T01, K | ATfl, K | $d_{01}$, Å | $\xi_c(0)$, Å | $C_{3D}$ |
|---|---|---|---|---|---|---|---|---|---|
| 99.8 | 6.4 | 6.62 | 6.67 | 6.8 | 9.4 | 2.8 | 4.08 | 2.67 ± 0.02 | 0.38 |

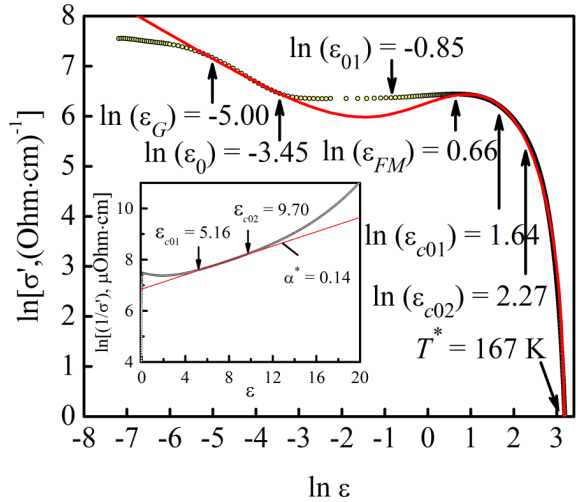

**FIG. 4.** Dependence of ln σ′ on ln ε in the Dy$_{0.6}$Y$_{0.4}$Rh$_{3.85}$Ru$_{0.15}$B$_4$ polycrystal in comparison with Eq. (9) (solid red curve). Insert: parameter definition of theory[51] ε*$_{c0}$ = 1 / α = 7.14 (see text).

$2\Delta(0)/k_BT_c = 2\Delta^*(T_G)/k_BT_c$ in a specific sample.[5,41,46] The best approximation of the ln σ′–1/T relationship by Eq. (9) for Dy$_{0.6}$Y$_{0.4}$Rh$_{3.85}$Ru$_{0.15}$B$_4$ is achieved at $2\Delta^*(T_G)/k_BT_c = 7.0 \pm 0.1$. This value of $2\Delta^*(T_G)/k_BT_c$ is typical for HTSC cuprates of Bi$_{1.6}$Pb$_{0.4}$Sr$_{1.8}$Ca$_{2.2}$Cu$_3$O$_x$ (Bi2223) ($T_c \approx 110$ K)[57] and Bi2212 with various $T_c$,[58] but is somewhat unexpected for Dy$_{0.6}$Y$_{0.4}$Rh$_{3.85}$Ru$_{0.15}$B$_4$ with $T_c = 6.4$ K. However, it is significant that the same value of $2\Delta(T_c)/k_BT_c = 7.2$ is obtained from the analysis of the Andreev spectra for Au–Dy$_{0.6}$Y$_{0.4}$Rh$_{3.85}$Ru$_{0.15}$B$_4$ contacts in a zero magnetic field at $T = 1.6$ K (see Fig 2 in Ref. 14. Lee us note that large values of $2\Delta_1(T_c)/k_BT_c \sim 9$ ($\Delta_1 \approx 3.5$ meV) and $2\Delta_2(T_c)/k_BT_c \sim 6.5$ ($\Delta_2 \approx 2.5$ meV) for FeSe single crystals with $T_c = 8.5$ K, according to the authors, indicate a very unusual mechanism of SC pairing in FeSe associated with the band structure specifics.[59] Thus, the large value of ratio $2\Delta(T_c)/k_BT_c \sim 7$ in combination with the relatively small value of $T_c$ and a large intrinsic magnetic moment of Dy indicates an unconventional (possibly triplet[10–15]) SC pairing mechanism in Dy$_{0.6}$Y$_{0.4}$Rh$_{3.85}$Ru$_{0.15}$B$_4$, which is different from the BCS mechanism.[3–6] The obtained result allows us to explain the relatively small value of $\xi_c(0) = (2.67 \pm 0.02)$ Å found in the experiment, which is typical for HTSCs with strong coupling.[5,17,32,51,56]

Since cuprates reveal an abnormally large energy gap $\Delta(0) = \Delta_0$, the ratio $2\Delta/k_BT_c \sim 7$ significantly exceeds the limit of the BCS theory for d-wave superconductors ($2\Delta/k_BT_c \sim 4.28$).[60,61] The large deviation of the $2\Delta/k_BT_c$ ratio from the BCS theory can be explained in the strong coupling theory,[62–64] if a decisive contribution to the pairing mechanism is made by delayed interactions with bosons of low-energy $\Omega_0$ which is comparable with the $\Delta_0$ parameter.[57] Among these theories, the most popular is the model in which Cooper pairing in HTSCs is realized as a result of the interaction between electrons and spin fluctuations.[65–67]

It is assumed that the so-called resonant spin mode makes a significant contribution,[68] which gives Cooper pairing a delayed strongly coupled nature[67,69,70] and makes it possible to explain the observed large ratio of $2\Delta/k_BT_c$.[57,58] Spin-fluctuation interaction leads to electron repulsion. However, if processes with high momentum transfer predominate in the spin fluctuation exchange, this may result in the formation of Cooper pairs with the d-wave symmetry of the order parameter.[65,67] In this case, $\Delta_0$ corresponds to the maximum value of the energy gap.

The experimental proof of the d-wave symmetry of the energy gap in cuprates (e.g. see Ref. 57 and references therein) served as a strong argument in favor of the spin-fluctuation HTSC model. However, recent findings of high angular resolution photoemission spectroscopy (ARPES),[71] and scanning tunneling spectroscopy[72–74] showed that the pairing mechanism in HTSCs can have a weakly coupled nature because the critical temperature $T_c$ is defined by $\Delta_{SC}$ which is significantly lower than $\Delta_0$. As a result, $2\Delta_{SC}/k_BT_c \sim 4.3$, which corresponds to the BCS theory for a d-wave superconductor.[60] In this case, low-frequency spin excitations underlying the spin-fluctuation model[62–64,68] are not critical. Therefore, despite the progress in the spectroscopy of bosonic excitations in cuprates,[61–64] it has not been possible, so far, to prove the effectiveness of the interaction of electrons with low-frequency bosonic modes, which could explain the observed large ratio of $2\Delta_0/k_BT_c$.[58,60,75] This conclusion, however, contradicts the results of microcontact spectroscopy (MCS),[58,75] as well as the conclusions of the theory,[76,77] from which it follows that $2\Delta/k_BT_c \sim 5$ for YBCO and $2\Delta/k_BT_c \sim 7$ for BiSCCO Similar results are obtained from the PG analysis in cuprates.[5,17,41,46] Thus, the question remains open.

Unfortunately, similar studies have not been carried out for Dy$_{0.6}$Y$_{0.4}$Rh$_{3.85}$Ru$_{0.15}$B$_4$ Therefore, the mechanism of the SC state implementation in these compounds is apparently even more complex, especially if we take into account the large intrinsic magnetic moment of Dy ions. A large value of $2\Delta^*/k_BT_c \sim 7$, which is not typical for such values of $T_c$, also speaks in favor of this conclusion. It can also be assumed that the formation of excess conductivity in HTSCs, including Dy$_{0.6}$Y$_{0.4}$Rh$_{3.85}$Ru$_{0.15}$B$_4$, corresponds precisely to $\Delta_0$, which explains the large values of $2\Delta/k_BT_c$ observed in these compounds. We assumed that the temperature dependence of the PG can provide an answer to some of the questions raised.

Solving Eq. (9) with respect to $\Delta^*(T)$, we obtain

$$\Delta^*(T) = T \ln \frac{e^2 A_4 (1 - T/T^*)}{\sigma'(T) 16\hbar \xi_c(0) \sqrt{2\varepsilon^*_{c0}} \text{sh}(2\varepsilon/\varepsilon^*_{c0})}, \quad (10)$$

where σ′(T) is the excess conductivity experimentally measured over the entire temperature range of $T^*$ to $T_c^{mf}$. The fact that σ′(T) is well described by Eq. (9) (Fig. 4) suggests that Eq. (10) yields a reliable magnitude and temperature dependence of $\Delta^*$. Figure 5 displays the analysis of $\Delta^*(T)$ according to (10) and using the following parameters defined from the experiment: $T_c^{mf} = 6.62$ K, $T^* = 167$ K, $\xi_c(0) = 2.67$ Å, $\varepsilon^*_{c0} = 7.14$, $A_4 = 11$ and $\Delta^*(T_G)/k_B = 22$ K. The obtained dependence is typical for magnetic HTSCs such as EuFeAsO$_{0.85}$F$_{0.15}$,[18] FeSe$_{0.94}$,[21] and, as can be seen, differs significantly from the similar dependence for non-magnetic cuprates.[17,26] The curve $\Delta^*(T)$ (Fig. 5) depicts a number of features that are observed at the corresponding characteristic temperatures. Thus, below $T^* = 167$ K, a pronounced maximum is observed at $T_{max} = 154$ K, which is typical of magnetic superconductors.[5,18] Then, a minimum follows at a temperature of $T_{min} \approx 95$ K. In FeSe compounds,[21,59] a similar minimum corresponds to a structural phase transition from a tetra- to an ortho-phase at $T_s \sim 90$ K, indicating the possibility of a similar structural transition in $Dy_{0.6}Y_{0.4}Rh_{3.85}Ru_{0.15}B_4$. Below $T_{min}$, $\Delta^*(T)$ increases, showing a wide maximum at $T_{pair} \approx 36$ K, followed by a minimum at $T_{01} = 9.4$ K. This behavior resembles the dependence $\Delta^*(T)$ for cuprates and indicates the possibility of implementing the PG state in the interval of $T < T_{min}$, as is assumed in FeSe at $T < T_s$.[78] To confirm this assumption, dependence $\Delta^*(T)$ in Fig. 6 is plotted in the temperature range of 0–100 K and 12–24 K along the Y axis. Dependence $\Delta^*(T)$ of this type, with a broad maximum at $T_{pair} \approx 36$ K and a pronounced minimum at $T_{01} = 9.4$ K, is typical of well-structured cuprates,[17,41] which confirms the assumption made. Thus, it can be expected that, below $T_{pair}$ in $Dy_{0.6}Y_{0.4}Rh_{3.85}Ru_{0.15}B_4$, fluctuation Cooper pairs (FCPs) begin to form, as in the case of cuprates.[17,27–29,44] Accordingly, below $T_{01}$, the system goes into the region of SC fluctuations, in which, as noted above, FCPs behave almost like SC pairs, but without long-range ordering (the so-called 'short-range phase correlations'). As a result, below $T_{01}$, dependence $\Delta^*(T)$ in $Dy_{0.6}Y_{0.4}Rh_{3.85}Ru_{0.15}B_4$ appears completely the same as in all other HTSCs: near $T_c$, as always, a maximum is observed at $T \sim T_0$ and a minimum at $T = T_G$ (see the insert in Fig. 5). Below $T_G$, there is a sharp jump of $\Delta^*(T)$ at $T \to T_c^{mf}$, but this is already a transition to the region of critical fluctuations, where the LP model does not work. Thus, the LP model makes it possible to determine the exact values of $T_G$ and, as a result, to obtain reliable values of $\Delta^*(T_c^{mf}) \approx \Delta^*(T_G) = \Delta(0) \approx 2$ meV and $2\Delta^*(T_c)/k_B T_c \approx 7$. Notably, on $\Delta^*(T)$, there is no sharply defined peculiarity at the magnetic transition temperature $T_{FM} = 19$ K, except that $\Delta^*(T)$ begins to decrease slightly more intensively at $T < 19$ K than is observed in FeSe.[21] However, strictly speaking, the magnetic maximum observed in Fig. 3 at $T_{FM}$ (ln $\varepsilon_{FM} = 0.66$), is no longer so noticeable in Fig. 4. That is, even on the ln $\sigma'$–ln $\varepsilon$ dependence, the peculiarity during the magnetic transition is very weakly expressed.

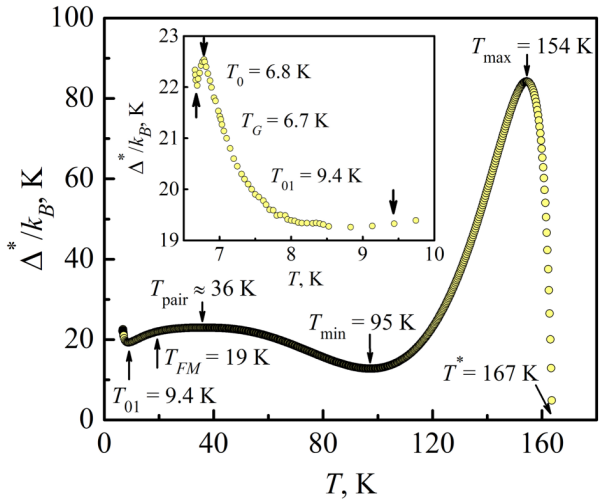

**FIG. 5.** Temperature dependence of the PG parameter $\Delta^*(T)/k_B$ for a $Dy_{0.6}Y_{0.4}Rh_{3.85}Ru_{0.15}B_4$ polycrystal. The insert shows $\Delta^*(T)/k_B$ in the region of superconducting fluctuations near $T_c$. All characteristic temperatures are arrowed.

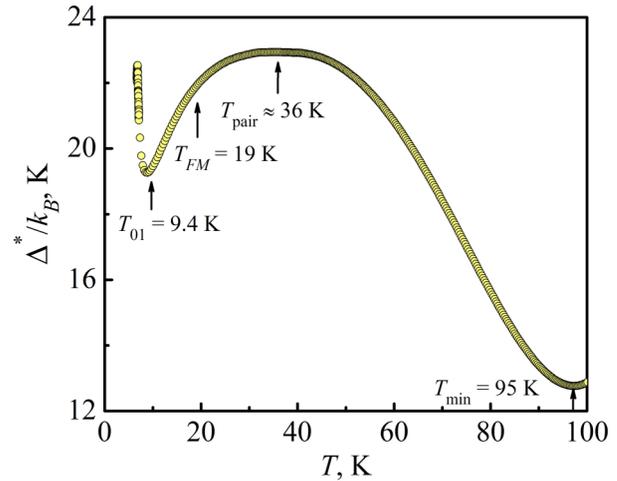

**FIG. 6.** Temperature dependence of the PG parameter $\Delta^*(T)/k_B$ of $Dy_{0.6}Y_{0.4}Rh_{3.85}Ru_{0.15}B_4$ near $T_c$.

At the same time, the form of dependence $\Delta^*(T)$ in $Dy_{0.6}Y_{0.4}Rh_{3.85}Ru_{0.15}B_4$ near $T_c$, with a maximum at $T \sim T_0$ and a minimum at $T = T_G$ (see the insert in Fig. 5) is, in fact, the same as the temperature dependence of the local pair density in HTSCs, $\langle n_\uparrow n_\downarrow \rangle$ calculated in the Peters–Bauer (PB) theory[22] in the framework of the 3D Hubbard model with attraction for various values of temperature $T/W$, interaction $U/W$ and the filling factor, where $W$ is the bandwidth. This allows us to compare the experimental values of $\Delta^*/\Delta^*_{max}$ with the PB theory and estimate the value of $\langle n_\uparrow n_\downarrow \rangle$ in $Dy_{0.6}Y_{0.4}Rh_{3.85}Ru_{0.15}B_4$ at $T_G$. For this purpose, let us bring the values of $\Delta^*/\Delta^*_{max}$ at $T_G$ in coincidence with a minimum, and at $T_0$ with the maximum of each theoretical curve calculated for various values of $U/W$. The fitting results for the three values of $U/W$ are shown in Fig. 7. It can be seen that the best agreement of the results, moreover in a wide range of $T/W$, 0 to 0.7, is observed at $U/W = 0.4$. Hence, $\langle n_\uparrow n_\downarrow \rangle$ ($T_G$) ≈ 0.35, which is noticeably greater than $\langle n_\uparrow n_\downarrow \rangle$ ($T_G$) ≈ 0.3 obtained for optimally doped YBaCuO single crystals.[38] This somewhat unexpected result can be explained by two reasons. Firstly, the strong intrinsic magnetism of Dy contributes to the growing number of PCFs. Here, it is assumed that the role of magnetism in the SC pairing mechanism in $Dy_{0.6}Y_{0.4}Rh_{3.85}Ru_{0.15}B_4$ is very prominent. Secondly, as discussed in the Introduction, there is a possibility of unconventional, e.g. triplet, pairing in the superconductors[8–11] whose strong magnetism coexists with superconductivity, which appears to be another contribution factor to the increase in $\langle n_\uparrow n_\downarrow \rangle$.

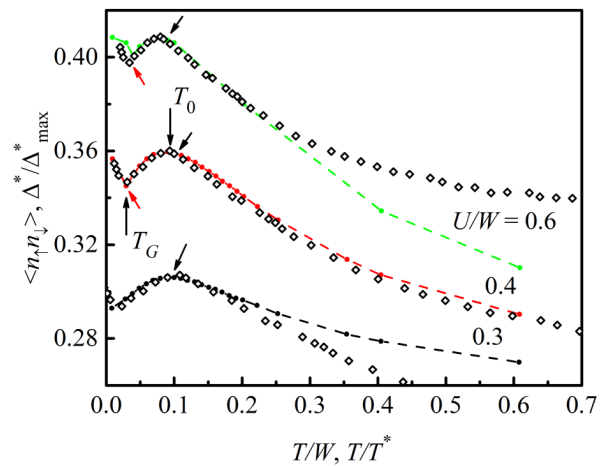

**FIG. 7.** Comparison of the experimental dependencies $\Delta^*/\Delta^*_{max}$ on $T/T^*$ (rhombi) of a $Dy_{0.6}Y_{0.4}Rh_{3.85}Ru_{0.15}B_4$ polycrystal with theoretical dependencies $\langle n_\uparrow n_\downarrow \rangle$ of $T/W$ for the three values of the interaction $U/W$: 0.3, 0.4 and 0.6.[22]

## 3. CONCLUSIONS

Temperature dependencies of the excess conductivity $\sigma'(T)$ and the possible pseudogap (PG), $\Delta^*(T)$, were first studied in the magnetic superconductor $Dy_{0.6}Y_{0.4}Rh_{3.85}Ru_{0.15}B_4$. It was shown that $\sigma'(T)$ near $T_c$ well described by the 3D Aslamazov–Larkin equation, demonstrating a 3D-2D crossover with increasing temperature. Using the crossover temperature $T_0$, the coherence length was measured along the $c$ axis, $\xi_c(0) = (2.67 \pm 0.02)$ Å, which correlates with the literature data for strong coupling HTSCs.[5,17,32,38,78] The pronounced effect of magnetism is found in the unusual dependence of $\ln \sigma'$ on $\ln \varepsilon$ with a maximum at $T_{FM} \sim 19$ K, which is associated with the system transition to the ferromagnetic state with decreasing temperature.

The dependence $\Delta^*(T)$ revealed a number of peculiarities typical of superconductors admit the possibility of the superconductivity–magnetism interplay. This is a high narrow maximum at $T = 154$ K, typical of magnetic superconductors, followed by a minimum at $T_{min} \approx 95$ K. In FeSe compounds, a similar minimum corresponds to the structural phase transition from the tetra- to the ortho-phase at $T_s \sim 90$ K,[21] indicating the possibility of a similar structural transition in $Dy_{0.6}Y_{0.4}Rh_{3.85}Ru_{0.15}B_4$. Below $T_{min}$, $\Delta^*(T)$ again increases, demonstrating a broad maximum at $T_{pair} \approx 36$ K, followed by a minimum at $T_{01} = 9.4$ K. This form of $\Delta^*(T)$ is similar to the temperature dependence of the pseudogap in cuprates,[17,26] which indicates the possibility of implementing the PG state in $Dy_{0.6}Y_{0.4}Rh_{3.85}Ru_{0.15}B_4$ at $T < T_{min}$, as is the case in FeSe at $T < T_s$.[78] It was shown that, below $T_{01}$, $\Delta^*(T)$ in $Dy_{0.6}Y_{0.4}Rh_{3.85}Ru_{0.15}B_4$ is the same as in all HTSCs with a maximum at $T \sim T_0$ and a minimum at $T = T_G$,[5,17,26] which indicates a common behavior of both magnetic and nonmagnetic superconductors in the region of superconducting fluctuations near $T_c$.

Meanwhile, the analysis of $\Delta^*(T)$ in $Dy_{0.6}Y_{0.4}Rh_{3.85}Ru_{0.15}B_4$ reveals a number of peculiarities. The first is an unexpectedly large value of $2\Delta^*(T_c)/k_BT_c = 7.0 \pm 0.1$. It is remarkable, however, that the same value of $2\Delta(T_c)/k_BT_c \sim 7.2$ is obtained from the Andreev spectral analysis of Au–$Dy_{0.6}Y_{0.4}Rh_{3.85}Ru_{0.15}B_4$ contacts measured in a zero magnetic field at $T = 1.6$ K.[14] This result indicates a more complicated mechanism of the SC state implementation in such superconductors as compared to cuprates, especially taking into account the large intrinsic magnetic moment of Dy ions. Secondly, it is the high density of local pairs $\langle n_\uparrow n_\downarrow \rangle$ obtained by comparing the experimental values of $\Delta^*/\Delta^*_{max}$ with the Peters–Bauer theory.[22] The measured $\langle n_\uparrow n_\downarrow \rangle$ ($T_G$) ~ 0.35 appears 1.17 times greater than $\langle n_\uparrow n_\downarrow \rangle$ ($T_G$) obtained for optimally doped YBaCuO single crystals.[38] This result can be explained by the fact that strong intrinsic magnetism of Dy can contribute to the increasing number of FCPs. In this regard, the role of magnetism in the SC pairing mechanism in $Dy_{0.6}Y_{0.4}Rh_{3.85}Ru_{0.15}B_4$ is assumed to be very important. Furthermore, as discussed in the Introduction, the possibility of unconventional, e.g. triplet, pairing in superconductors[8–12] whose strong magnetism coexists with superconductivity can also lead to the increase in $\langle n_\uparrow n_\downarrow \rangle$.

Translated by AIP Author Services